\documentclass[entropy,article,accept,moreauthors,pdftex]{Definitions/mdpi}

\usepackage{amsmath}
\usepackage{amsthm}
\usepackage{array}
\usepackage{fancybox}
\usepackage{multirow}
\usepackage{appendix}
\usepackage{epsfig}
\usepackage{cases}
\usepackage{forest}
\usepackage{color}
\usepackage[labelformat=simple]{subcaption}

\DeclareCaptionLabelFormat{subcaptionlabel}{\normalfont(\textbf{#2}\normalfont)}
\firstpage{1} 
\pubvolume{21}
\issuenum{1}
\articlenumber{280}
\pubyear{2019}
\copyrightyear{2019}
\history{Published in {\it{Entropy}} {\bf{2019}}, 21(3), 280; \url{https://doi.org/10.3390/e21030280}}
\updates{yes} 

\setitemize{parsep=6pt,itemsep=0pt,leftmargin=*,labelsep=5.5mm}
\setenumerate{parsep=6pt,itemsep=0pt,leftmargin=*,labelsep=5.5mm}
\setlist[description]{itemsep=0mm}

\usepackage{environ}
\NewEnviron{myequation}{%
\begin{equation}
\scalebox{0.88}{$\BODY$}
\end{equation}}

\Title{Non-Geodesic Incompleteness in Poincar\'e Gauge~Gravity}

\Author{J. A. R. Cembranos $^{1,}$*\orcidA{}, J. Gigante Valcarcel $^{1}$\orcidB{} and F. J. Maldonado Torralba $^{2,3}$\orcidC{}}

\AuthorNames{J. A. R. Cembranos, J. Gigante Valcarcel and F.J. Maldonado Torralba}

\address{
$^{1}$ \quad Departamento de F\'{\i}sica Te\'orica and IPARCOS, Universidad Complutense de Madrid, E-28040 Madrid, Spain; jorgegigante@ucm.es\\
$^{2}$ \quad Cosmology and Gravity Group, Department of Mathematics and Applied Mathematics, University of Cape Town, Rondebosch 7701, Cape Town, South Africa; f.j.maldonado.torralba@rug.nl\\
$^{3}$ \quad Van Swinderen Institute, University of Groningen, 9747 AG Groningen, The Netherlands}

\corres{Correspondence: cembra@fis.ucm.es}

\abstract{In this work, we review the study of singularities in Poincar\'e gauge theories of gravity. Since one of the most recent studies used the appearance of black hole regions of arbitrary dimension as an indicator of singular behavior, we also give some explicit examples of these structures and study how particles behave around them.}


\begin{document}

\maketitle

\section{Introduction}

\emph{Can spin avert singularities?} This is a question that has been under study since Stewart and Hajicek proposed that the presence of torsion, sourced by half-spin particles, would lead to the avoidance of singularities in the spacetime~\cite{Steward}. Before exploring the different answers to this question, let us review the concept of singularity in General Relativity (GR).

In a physical theory, a singularity is usually known as a ``place'' where some of the quantities used in the description of a dynamical system diverge. As an example, we can find this situation evaluating the Coulomb potential $V=K\frac{q}{r}$ at the point $r=0$. This kind of behavior appears because the theory is not valid in the considered region or we have assumed a simplification. Namely, in the previous example, the singularity arises due to the fact that we are considering the charged particle as a point and omitting the quantum effects.

In GR, one might expect to observe singularities when certain components of the tensors that describe the spacetime geometry diverge. This means that curvature is higher than $\frac{1}{l_{p}^{2}}$, where $l_{p}$ is the Planck length, so we need to take into account the quantum effects, which are not considered in this theory. However, there are situations where this behavior is given as a result of the election of the coordinate system. This is the case of the ``singularity'' at $r=2M$ in the Schwarzschild solution. For this reason, another criteria, proposed by Penrose~\cite{Penrose}, is used to define a spacetime singularity: geodesic incompleteness. The physical interpretation of this condition is the existence of free-falling observers that appear or disappear out of nothing. This is ``strange'' enough to consider it as a sufficient condition to ensure the occurrence of singularities.

In general, all singularity theorems follow the same pattern, made explicit by Senovilla in~\cite{Senovilla}:
\begin{Theorem}
(Pattern singularity ``theorem'') If the spacetime satisfies:
\begin{enumerate} [leftmargin=2.3em,labelsep=4mm]
\item[(1)] A condition on the curvature tensor. 
\item[(2)] A causality condition. 
\item[(3)] An appropriate initial and/or boundary condition. 
\end{enumerate}
Then, there are null or timelike inextensible incomplete geodesics. 
\end{Theorem}

The reasoning that one can follow to answer if the introduction of new geometrical degrees of freedom can avoid the appearance of singularities is to study if the standard conditions exposed in this theorem may change in the framework under consideration. In this sense, the authors proved in~\cite{JJFJ} that in a strongly asymptotically-predictable spacetime, the conditions for having a singular trajectory for any massive particle in theories with torsion are the same as in GR. 

The conditions of the singularity theorem established in~\cite{JJFJ} predict the existence of black hole (BH) regions of arbitrary dimension. This type of structure has been under study in torsion theories since the beginning of the 1980s, either finding gravitational models where the Birkhoff theorem holds~\cite{Nev,Rauch-Nieh,delaCruz-Dombriz:2018vzn} or finding new BH solutions sourced by dynamical torsion~\cite{Bakler:1988nq,Obukhov:1987tz,JJ,JJJ,Shabani:2017owe}.

The purpose of this work is to provide more insights into the singularity problem in torsion theories. For this reason, we have organized it as follows. First, in Section \ref{geometry}, we present the general geometric structure of these kinds of theories. In Section \ref{singularity}, we review the singularity theorem for strongly asymptotically-predictable spacetimes in the presence of torsion, remarking on its physical consequences. In Sections \ref{BH} and \ref{fermion}, we focus on a BH region with a Reissner--Nordstr\"{o}m-like geometry sourced by dynamical torsion, in order to analyze the behavior of Dirac particles minimally coupled to torsion (i.e., coupled to the axial component of torsion), giving an explicit example of the application of the singularity theorem. Finally, we expose our conclusions in Section \ref{conclusions}.

\section{Generalities of Theories with Torsion}
\label{geometry}

The geometric foundations of gravitational theories characterized by a metric-compatible asymmetric connection can be naturally systematized as a gauge theory of the Poincar\'e group~\cite{Hehl,Gauge}. The model requires gauging the external degrees of freedom consisting of rotations and translations, which means that a gauge connection containing two principal independent variables can be introduced in order to describe the gravitational field:\begin{equation}A_{\mu}=e^{a}\,_{\mu}P_{a}+\omega^{a b}\,_{\mu}J_{a b}\,,\end{equation}
where $e^{a}\,_{\mu}$ is the vierbein field and $\omega^{a b}\,_{\mu}$ is the spin connection, satisfying the relations~\cite{Yepez}:\begin{equation}
g_{\mu \nu}=e^{a}\,_{\mu}\,e^{b}\,_{\nu}\,\eta_{a b}\,,
\end{equation}\begin{equation}
\omega^{a b}\,_{\mu}=e^{a}\,_{\lambda}\,e^{b \rho}\,\Gamma^{\lambda}\,_{\rho \mu}+e^{a}\,_{\lambda}\,\partial_{\mu}\,e^{b \lambda}\,.
\end{equation}

Thereby, the vierbein field and the affine connection act as translational- and rotational-type potentials. Since the connection is not necessarily symmetric, we must deal with a Riemann--Cartan (RC) manifold endowed with curvature and torsion, which involves the following relation between the general affine connection and the Levi--Civita (LC) connection:\begin{equation}
\label{eq2}
\mathring{\Gamma}^{\rho}\,_{\mu\nu}=\Gamma^{\rho}\,_{\mu\nu}-K^{\rho}\,_{\mu\nu}\,,
\end{equation}
where:\begin{equation}
K^{\rho}\,_{\mu\nu}=\frac{1}{2}\left(T^{\rho}\,_{\mu\nu}-T_{\mu}\,^{\rho}\,_{\nu}-T_{\nu}\,^{\rho}\,_{\mu}\right)\,,
\end{equation}
is the \emph{contortion} tensor. In addition, the curvature tensors depend on the relative affine connection, so~that it is essential to distinguish between the torsion-free Riemann curvature and the RC curvature depending on torsion~\cite{Sabbata,Shapiro}:
\begin{equation}
\label{eq10}
\mathring{R}^{\lambda}\,_{\mu\nu\rho}=R^{\lambda}\,_{\mu\nu\rho}-\mathring{\nabla}_{\nu}K^{\lambda}\,_{\mu\rho}+\mathring{\nabla}_{\rho}K^{\lambda}\,_{\mu\nu}-K^{\lambda}\,_{\sigma\nu}K^{\sigma}\,_{\mu\rho}+K^{\lambda}\,_{\sigma\rho}K^{\sigma}\,_{\mu\nu}\,,
\end{equation}
where the upper index $\mathring{\,}$ denotes torsion-free quantities.

By contraction, we can obtain the expression for the Ricci tensor:
\begin{equation}
\label{eq11}
\mathring{R}_{\mu\rho}=R_{\mu\rho}-\mathring{\nabla}_{\lambda}K^{\lambda}\,_{\mu\rho}+\mathring{\nabla}_{\rho}K^{\lambda}\,_{\mu\lambda}-K^{\lambda}\,_{\sigma\lambda}K^{\sigma}\,_{\mu\rho}+K^{\lambda}\,_{\sigma\rho}K^{\sigma}\,_{\mu\lambda}\,,
\end{equation}
and for the scalar curvature:\begin{equation}
\mathring{R}=g^{\mu\rho}\mathring{R}_{\mu\rho}=R+2\mathring{\nabla}_{\lambda}K^{\rho\lambda}\,_{\rho}-K^{\lambda}\,_{\sigma\lambda}K^{\sigma\rho}\,_{\rho}+K^{\sigma}\,_{\lambda\rho}K^{\lambda\rho}\,_{\sigma}\,.
\end{equation}

By following these lines, it is straightforward to derive the fundamental relation between torsion and the corresponding field strength tensor of the translation group:\begin{equation}F^{a}\,_{\mu \nu}=e^{a}\,_{\lambda}\,T^{\lambda}\,_{\nu \mu}\,,\end{equation}
and between curvature and the field strength tensor of the rotation group:\begin{equation}
F^{a b}\,_{\mu \nu}=e^{a}\,_{\lambda}e^{b}\,_{\rho}\,R^{\lambda \rho}\,_{\mu \nu}\,.
\end{equation}

The introduction into the gravitational action of higher order corrections depending on the gauge curvatures extends the conventional approach and establishes a correspondence between the geometry of the spacetime and the energy-momentum and spin density tensors of matter. In fact, the presence of the spin density tensor implies the symmetric energy-momentum tensor of GR to be replaced by a canonical energy-momentum tensor, generally endowed with a non-vanishing antisymmetric component, which operates as a source of gravity and completes the fundamental relation between geometry and physics.

\section{The Singularity Theorem}
\label{singularity}

In order to review the singularity theorem proven by the authors in~\cite{JJFJ}, we need to introduce some definitions. We know intuitively that the existence of incomplete null geodesics usually leads to the appearance of BHs, which are regions of the spacetime beyond which an inside observer cannot escape. This applies to all timelike and null curves, not just geodesics. This is known as the \emph{cosmic censorship conjecture}, which was introduced by Penrose in 1969. It basically states that singularities cannot be \emph{naked}, meaning that they cannot be seen by an outside observer. However, how can this concept be expressed mathematically? The answer lies in the concept of \emph{conformal compactification}, which can be defined as~\cite{Infinity}:
\begin{Definition}
Let $\left(M,g\right)$ and $\left(\tilde{M},\,\tilde{g}\right)$ be two spacetimes. Then, $\left(\tilde{M},\,\tilde{g}\right)$ is said to be a conformal compactification of $M$ if and only if the following properties are met:
\begin{enumerate}
\item $M$ is an open submanifold of $\tilde{M}$ with smooth boundary $\partial\tilde{M}=\mathcal{J}$. This boundary is usually denoted \emph{conformal infinity}.
\item There exists a smooth scalar field $\Omega$ on $\tilde{M}$, such that $\tilde{g}_{\mu\nu}=\Omega^{2}g_{\mu\nu}$ on $M$ and so that $\Omega=0$ and its gradient $d\Omega\neq 0$ on $\mathcal{J}$.
\end{enumerate}
If additionally, every null geodesic in $M$ acquires a future and a past endpoint on $\mathcal{J}$, the spacetime is called \emph{asymptotically simple}. Furthermore, if the Ricci tensor is zero in a neighborhood of $\mathcal{J}$, the spacetime is said to be \emph{asymptotically empty}.
\end{Definition} 
In a conformal compactification, $\mathcal{J}$ is composed of two null hypersurfaces, $\mathcal{J}^{+}$ and $\mathcal{J}^{-}$, known as \emph{future null infinity} and \emph{past null infinity}, respectively.

In order to establish the definition of a BH, we need to introduce two additional concepts~\cite{Wald}:
\begin{Definition}
A spacetime $\left(M,g\right)$ is said to be \emph{asymptotically flat} if there is an asymptotically-empty spacetime $\left(M',g'\right)$ and a neighborhood $\mathcal{U}'$ of $\mathcal{J}'$, such that $\mathcal{U}'\cap M'$ is isometric to an open set $\mathcal{U}$ of $M$. 
\end{Definition} 
\begin{Definition}
Let $\left(M,g\right)$ be an asymptotically-flat spacetime with conformal compactification $\left(\tilde{M},\,\tilde{g}\right)$. Then,~$M$ is called \emph{(future) strongly asymptotically predictable} if there is an open region $\tilde{V}\subset\tilde{M}$, with~$\overline{J^{-}\left(\mathcal{J}^{+}\right)\cap M}\subset\tilde{V}$, such that $\tilde{V}$ is globally hyperbolic. 
\end{Definition}
This definition does not require the condition of endpoints of the null geodesics, meaning that these types of spacetimes can be singular. Nevertheless, if a spacetime is asymptotically predictable, then the singularities are not naked (i.e., are not visible from $\mathcal{J}^{+}$).

Now, we can establish what we understand by a BH:
\begin{Definition}
A strongly asymptotically-predictable spacetime $\left(M,g\right)$ is said to contain a BH if $M$ is not contained in $J^{-}\left(\mathcal{J}^{+}\right)$. The BH region, $B$, is defined to be $B=M-J^{-}\left(\mathcal{J}^{+}\right)$, and its boundary, $\partial B$, is known as the \emph{event horizon}. 
\end{Definition}
Intuitively, we think that a particle in a closed trapped surface cannot escape to $\mathcal{J}^{+}$, meaning that it is part of the BH region of the spacetime. Nevertheless, this is not true in general. In the next proposition, we establish the conditions that ensure the existence of BHs when we have a closed future trapped submanifold of arbitrary co-dimension:
\begin{Proposition}
Let $\left(M,g\right)$ be a strongly asymptotically-predictable spacetime of dimension $n$ and $\Sigma$ a closed future trapped submanifold of arbitrary co-dimension $m$ in $M$. If the curvature condition holds along every future-directed null geodesic emanating orthogonally from $\Sigma$, then $\Sigma$ cannot intersect $J^{-}\left(\mathcal{J}^{+}\right)$ (i.e., $\Sigma$ is in the BH region $B$ of $M$\footnote{Analogously, it can be defined as a past strongly asymptotically-predictable spacetime, and~then, the proposition would predict the existence of white hole (WH) regions, $B=M-J^{+}\left(\mathcal{J}^{-}\right)$, which are regions where particles cannot enter, only exit.})
.
\end{Proposition}
\begin{proof}
The proof can be found in~\cite{JJFJ}.
\end{proof}

Now, the reader might be wondering how this proposition is related to the actual singularities of~particles.

From the minimal coupling procedure, it follows that particles without spin, represented by scalar fields, do not feel torsion, due to the fact that the covariant derivative of a scalar field is just its partial derivative. Furthermore, it is impossible to perform the minimal coupling prescription for Maxwell's field preserving the $U\left(1\right)$ gauge invariance; the Maxwell equations are the same as the ones present in GR. Therefore, they move following null extremal curves, so that the causal structure is determined by the metric structure, just like in GR. This means that the usual test particles follow the geodesic curves provided by the LC
 connection, which allows us to generalize the singularity theorems. However, what~happens when we consider fermions coupled to the spacetime torsion?
 
All the analyses of the trajectories that follow these kinds of particles (for a nice review, see~\cite{ProbeB}) have one thing in common: they experiment with a corrective factor of the form:\begin{equation}
\label{spineq}
a^{\mu}=v^{\rho}\mathring{\nabla}_{\rho}v_{\mu}=C\left(\frac{\hbar}{m}\right)f\left({R}_{\rho\sigma\lambda}\,^{\mu}s^{\rho\sigma}v^{\lambda}+K_{\sigma\rho}\,^{\mu}p^{\rho}v^{\sigma}\right),
\end{equation}
where $C$ is a constant, $m$ is the mass of the particle, and $s^{\rho\sigma}$ describes the internal spin tensor, related to the spin $s^{\mu}$ of the particle by:\begin{equation}
s^{\mu}=\frac{1}{2}\epsilon^{\mu\nu\rho\sigma}v_{\nu}s_{\rho\sigma},
\end{equation}
with $\epsilon^{\mu\nu\rho\sigma}$ the totally antisymmetric LC tensor.

It is clear from the previous analysis that massive spinning particles do not follow timelike geodesics. Nevertheless, independently of how torsion affects these particles, they will follow timelike curves, and we assume that locally (in a normal neighborhood of a point), nothing can be faster than light (null geodesics). Hence, it would be interesting to see under which circumstances we have incompleteness of non-geodesic timelike curves. For this reason, we recover the definition of an n-dimensional BH and WH given in Section \ref{singularity}. From this definition, we conclude that if these kinds of structures exist in our spacetime, we would have timelike curves (including non-geodesics) that do not have endpoints in the conformal infinity, since for the case of BHs, the spacetime $M$ is not contained in $J^{-}\left(\mathcal{J}^{+}\right)$, while for WHs, $M$ is not contained in $J^{+}\left(\mathcal{J}^{-}\right)$. Considering these lines, we establish the following theorem:

\begin{Theorem}
\label{thmsing}
Let $\left(M,g\right)$ be a strongly asymptotically-predictable spacetime of dimension $n$ and $\Sigma$ a closed future trapped submanifold of arbitrary co-dimension $m$ in $M$. If the curvature condition holds along every future-directed null geodesic emanating orthogonally from $\Sigma$, then some timelike curves in $M$ would not have endpoints in the conformal infinity; hence, $M$ is a singular spacetime.
\end{Theorem} 

One might wonder if one of the incomplete timelike curves actually represents the trajectory of a spinning particle coupled to the torsion tensor. From Equation (\ref{spineq}), which represents the non-geodesic behavior, we see that the only possible way that all the trajectories have endpoints in the conformal infinity is the existence of huge values for the curvature and torsion tensors near the event horizon, which in a physically-plausible scenario is not possible. This is why we consider it a more physically-relevant theorem for the singular behavior of such particles, since it is strongly related to the actual trajectories. 

With this reasoning, we have shown that in strongly-asymptotically predictable spacetimes, one cannot avoid the occurrence of singularities, even in the presence of torsion, if the conditions of Theorem \ref{thmsing} hold. {On the opposite case, it is possible to find non-singular configurations if the conditions are violated~\cite{Poplawski:2012ab,Lucat:2015rla}. }

In this sense, in the following section, we present an explicit example of a BH region sourced by the torsion field.

\section{Reissner--Nordstr\"{o}m Sourced by Torsion}
\label{BH}

In the framework of the Poincar\'e gauge (PG) theory, the propagating character of torsion demands the presence of higher order curvature terms in the gravitational action. This feature represents a deep aspect in the nature of torsion, which may produce significant effects in the geometry of the spacetime even in the absence of matter sources.

In order to analyze the possible implications derived by the presence of a dynamical torsion beyond the standard approach of GR, we introduce the following gravitational action:\begin{myequation}
\label{action}
S=\frac{1}{64\pi}\int d^{4}x\sqrt{-g}\left[-4\mathring{R}+3\left(6c_{1}+c_{2}\right)R_{\lambda[\rho\mu\nu]}R^{\lambda[\rho\mu\nu]}+9\left(2c_{1}+c_{2}\right)R_{\lambda[\rho\mu\nu]}R^{\mu[\lambda\nu\rho]}+8d_{1}R_{[\mu\nu]}R^{[\mu\nu]}\right]\,.
\end{myequation}

In the present case, the respective geometric corrections are mediated by a massless torsion field, which in fact yields a non-vanishing metric curvature described in a first approximation by Einstein's model\footnote{The Lagrangian coefficients can be chosen to obtain different gravitational theories (for some criteria on the election of a large class of these PG theories, see~\cite{Obukhov:1987tz,Chen:1988mz,Sezgin,Kuhfuss,Cembra})}.

The corresponding field equations can be easily derived by performing variations with respect to the gauge potentials:\begin{equation}
\label{eq6}
\mathring{G}_{\mu}\,^{\nu}=2c_{1}T1_{\mu}\,^{\nu}+c_{2}T2_{\mu}\,^{\nu}
-\left(2c_{1}+c_{2}\right)T3_{\mu}\,^{\nu}+d_{1}\left(H1_{\mu}\,^{\nu}-H2_{\mu}\,^{\nu}\right)\,,
\end{equation}\begin{equation}
2c_{1}C1_{\left[\mu\lambda\right]}\,^{\nu}-c_{2}C2_{\left[\mu\lambda\right]}\,^{\nu}+\left(2c_{1}+c_{2}\right)C3_{\left[\mu\lambda\right]}\,^{\nu}-d_{1}\left(Y1_{\left[\mu\lambda\right]}\,^{\nu}-Y2_{\left[\mu\lambda\right]}\,^{\nu}\right)=0\,,
\end{equation}
where the tensors above are geometrical quantities defined in the following way:
\begin{equation}
\begin{array}{c}
\,\\
T1_{\mu}\,^{\nu}\equiv R_{\lambda\rho\mu\sigma}R^{\lambda\rho\nu\sigma}-\frac{1}{4}\delta_{\mu}\,^{\nu}R_{\lambda\rho\alpha\sigma}R^{\lambda\rho\alpha\sigma}\,,\\
\,\\
T2_{\mu}\,^{\nu}\equiv R_{\lambda\rho\mu\sigma}R^{\lambda\nu\rho\sigma}+R_{\lambda\rho\sigma\mu}R^{\lambda\sigma\rho\nu}-\frac{1}{2}\delta_{\mu}\,^{\nu}R_{\lambda\rho\alpha\sigma}R^{\lambda\alpha\rho\sigma}\,,\\
\\
T3_{\mu}\,^{\nu}\equiv R_{\lambda\rho\mu\sigma}R^{\nu\sigma\lambda\rho}-\frac{1}{4}\delta_{\mu}\,^{\nu}R_{\lambda\rho\alpha\sigma}R^{\alpha\sigma\lambda\rho}\,,\\
\,\\
H1_{\mu}\,^{\nu}\equiv R^{\nu}\,_{\lambda\mu\rho}R^{\lambda\rho}+R_{\lambda\mu}R^{\lambda\nu}-\frac{1}{2}\delta_{\mu}\,^{\nu}R_{\lambda\rho}R^{\lambda\rho}\,,\\
\,\\
H2_{\mu}\,^{\nu}\equiv R^{\nu}\,_{\lambda\mu\rho}R^{\rho\lambda}+R_{\lambda\mu}R^{\nu\lambda}-\frac{1}{2}\delta_{\mu}\,^{\nu}R_{\lambda\rho}R^{\rho\lambda}\,,\\
\,\\
C1_{\mu}\,^{\lambda\nu}\equiv\mathring{\nabla}_{\rho}R_{\mu}\,^{\lambda\rho\nu}+K^{\lambda}\,_{\sigma\rho}R_{\mu}\,^{\sigma\rho\nu}-K^{\sigma}\,_{\mu\rho}R_{\sigma}\,^{\lambda\rho\nu}\,,\\
\,\\
C2_{\mu}\,^{\lambda\nu}\equiv\mathring{\nabla}_{\rho}\left(R_{\mu}\,^{\nu\lambda\rho}-R_{\mu}\,^{\rho\lambda\nu}\right)+K^{\lambda}\,_{\sigma\rho}\left(R_{\mu}\,^{\nu\sigma\rho}-R_{\mu}\,^{\rho\sigma\nu}\right)-K^{\sigma}\,_{\mu\rho}\left(R_{\sigma}\,^{\nu\lambda\rho}-R_{\sigma}\,^{\rho\lambda\nu}\right)\,,\\
\,\\
C3_{\mu}\,^{\lambda\nu}\equiv\mathring{\nabla}_{\rho}R^{\rho\nu\lambda}\,_{\mu}+K^{\lambda}\,_{\sigma\rho}R^{\rho\nu\sigma}\,_{\mu}-K^{\sigma}\,_{\mu\rho}R^{\rho\nu\lambda}\,_{\sigma}\,,\\
\,\\
Y1_{\mu}\,^{\lambda\nu}\equiv\delta_{\mu}\,^{\nu}\mathring{\nabla}_{\rho}R^{\lambda\rho}-\mathring{\nabla}_{\mu}R^{\lambda\nu}+\delta_{\mu}\,^{\nu}K^{\lambda}\,_{\sigma\rho}R^{\sigma\rho}+K^{\rho}\,_{\mu\rho}R^{\lambda\nu}-K^{\nu}\,_{\mu\rho}R^{\lambda\rho}-K^{\lambda}\,_{\rho\mu}R^{\rho\nu}\,,\\
\,\\
Y2_{\mu}\,^{\lambda\nu}\equiv\delta_{\mu}\,^{\nu}\mathring{\nabla}_{\rho}R^{\rho\lambda}-\mathring{\nabla}_{\mu}R^{\nu\lambda}+\delta_{\mu}\,^{\nu}K^{\lambda}\,_{\sigma\rho}R^{\rho\sigma}+K^{\rho}\,_{\mu\rho}R^{\nu\lambda}-K^{\nu}\,_{\mu\rho}R^{\rho\lambda}-K^{\lambda}\,_{\rho\mu}R^{\nu\rho}\,.
\end{array}
\end{equation}

It is straightforward to check that this set of equations reduces to the regular equations of GR when the dynamical character of the torsion tensor vanishes, by virtue of the condition:\begin{equation}
\nabla_{\left[\mu\right.}T^{\rho}\,_{\left.\nu\sigma\right]}+T^{\lambda}\,_{\left[\mu\nu\right.}T^{\rho}\,_{\left.\sigma\right]\lambda}=0\,.
\end{equation}

On the other hand, in the dynamical regime, it is possible to find the following static and spherically-symmetric BH solution with a Reissner--Nordstr\"{o}m-like geometry characterized by a spin charge, when the Lagrangian coefficients satisfy the relations $c_{1}=-\,d_{1}/4$ and $c_{2}=-\,d_{1}/2$~\cite{JJ}:\begin{equation}
\begin{cases}
\label{RNsolution}
f\left(r\right) \equiv g_{tt} = -\,\frac{1}{g_{rr}} = 1-\frac{2m}{r}+\frac{d_{1}\kappa^{2}}{r^{2}} \;\; , \;\; g_{\theta_{1}\theta_{1}} = -\,r^{2} \;\; , \;\; g_{\theta_{2}\theta_{2}} = -\,r^{2}\sin^{2}\theta_{1} \;\; , \\
\,\\
T^{t}\,_{t r}=\frac{\dot{f}(r)}{2f(r)} \;\; , \;\;
T^{r}\,_{t r}=\frac{\dot{f}(r)}{2} \;\; , \;\;
T^{\theta_{k}}\,_{t \theta_{k}}=\frac{f(r)}{2r} \;\; , \;\;
T^{\theta_{k}}\,_{r \theta_{k}}=-\,\frac{1}{2r} \;\; ,\\
\,\\
T^{\theta_{k}}\,_{t \theta_{l}}=\frac{\kappa}{r}\,\varepsilon_{a b}\,e^{a \theta_{k}}\,e^{b}\,_{\theta_{l}}\, \;\; , \;\;
T^{\theta_{k}}\,_{r \theta_{l}}=-\,\frac{\kappa}{rf\left(r\right)}\,\varepsilon_{a b}\,e^{a \theta_{k}}\,e^{b}\,_{\theta_{l}}\, \;\; ;
\end{cases}
\end{equation}
where $k,l=1,2$ with $k \neq l$, $\varepsilon_{a b}$ is the two-dimensional LC symbol, and the dot $\dot{\,}$ denotes the derivative with respect to the radial coordinate.

Thereby, within this model, the axial component of the torsion tensor acts as a Coulomb-like potential depending on $\kappa$, in such a way that the conventional approach of GR is completely recovered when it vanishes. Furthermore, this type of configuration can also be extended to the case where the mentioned mode is considered massive~\cite{JJJ}: \begin{myequation}
S=\frac{1}{64\pi}\int d^{4}x\sqrt{-g}\left[-4\mathring{R}-d_{1}\left(6R_{\lambda[\rho\mu\nu]}R^{\lambda[\rho\mu\nu]}+9R_{\lambda[\rho\mu\nu]}R^{\mu[\lambda\nu\rho]}-8R_{[\mu\nu]}R^{[\mu\nu]}-3w\,T_{[\lambda\mu\nu]}T^{[\lambda\mu\nu]}\right)\right]\,.
\end{myequation}

In that case, the following Proca-like equation is additionally satisfied by the curvature and torsion~components:\begin{equation}
\mathring{\nabla}_{\lambda}R^{\lambda}\,_{[\rho \mu \nu]}-w\,T_{[\rho\mu\nu]}=0\,,
\end{equation}
where the coefficient $w$ is related to the mass of the axial component of torsion.

The existing correspondence between spin and torsion may involve significant effects on the behavior of those spinning particles coupled to this geometric quantity, especially within post-Riemannian geometries induced by a dynamical axial mode, since this is the unique component implicated in the interaction with Dirac fields, according to the minimal coupling principle~\cite{Sabbata}. This~means that we can focus on the present BH solution to analyze the motion of spin 1/2 particles minimally coupled to torsion and therefore to see how the singularity theorem actually holds.

\section{Fermion Dynamics}
\label{fermion}

To calculate how half-spin particles behave in a theory with torsion, it is possible to apply the WKB
 expansion of the Dirac equation~\cite{WKB}. Within this approximation, the evolution of the spinor field and the respective non-geodesic acceleration at first order of $\hbar$ are given by:\begin{equation}
v^{\mu}\widetilde{\nabla}_{\mu}b_{0}=0\,,
\end{equation}\begin{equation}
\label{eq:16}
a_{\mu}=v^{\varepsilon}\mathring{\nabla}_{\varepsilon}v_{\mu}=\frac{\hbar}{4m_{s}}\widetilde{R}_{\lambda\rho\mu\nu}\overline{b}_{0}\sigma^{\lambda\rho}b_{0}v^{\nu}\,,
\end{equation}
where $m_{s}$ represents the mass of the spinning particle, $b_{0}$ the normalized spinor of the first coefficient of the WKB expansion, $\sigma^{\lambda\rho}$ the corresponding spin matrices, and $\widetilde{R}_{\lambda\rho\mu\nu}$ the intrinsic part of the RC tensor associated with the totally antisymmetric component of the torsion tensor:\begin{equation}[\widetilde{\nabla}_{\mu},\widetilde{\nabla}_{\nu}]\,v^{\lambda}=\widetilde{R}^{\lambda}\,_{\rho \mu \nu}\,v^{\rho}+3\,g^{\sigma\rho}\,T_{\left[\rho\mu\nu\right]}\,\widetilde{\nabla}_{\sigma}v^{\lambda}\,,
\end{equation}
with:\begin{equation}
\widetilde{R}^{\lambda}\,_{\rho \mu \nu}=\partial_{\mu}\widetilde{\varGamma}^{\lambda}\,_{\rho \nu}-\partial_{\nu}\widetilde{\varGamma}^{\lambda}\,_{\rho \mu}+\widetilde{\varGamma}^{\lambda}\,_{\sigma \mu}\widetilde{\varGamma}^{\sigma}\,_{\rho \nu}-\widetilde{\varGamma}^{\lambda}\,_{\sigma \nu}\widetilde{\varGamma}^{\sigma}\,_{\rho \mu}\,,
\end{equation}\begin{equation}
\label{eq:17}
\widetilde{\varGamma}^{\lambda}\,_{\mu\nu}=\mathring{\Gamma}^{\lambda}\,_{\mu\nu}+\frac{3}{2}\,g^{\lambda\rho}\,T_{\left[\rho\mu\nu\right]}\,.
\end{equation}

Thus, we can calculate the acceleration related to the trajectories of Dirac particles within the geometry provided by the solution \eqref{RNsolution} and perform a numerical analysis to stress interesting differences with respect to the geodesic motion of GR~\cite{JJFJ2}. For this purpose, two conditions must be considered, in order to maintain the semiclassical approximation and the positive energy associated with the spinor state:\begin{equation}
\label{via}
\left(\overline{b}_{0}\sigma^{r\beta}b_{0}\right)v_{\beta}=0\,,
\end{equation}\begin{equation}
\dot{f} \left(r\right)\ll f\left(r\right)\,.
\end{equation}
Specifically, the first expression represents the radial component of the Pirani condition, which can be introduced as a supplementary constraint to allow the propagating equations to be solved and to ensure the conservation of mass along the final trajectory~\cite{Pirani}. On the other hand, the second one is a purely metric condition that comes from the torsion-free Riemann tensor, and it establishes the derivative of $f\left(r\right)$ to be at least two orders of magnitude below the value of $f\left(r\right)$. Thereby, this relation is a consequence of the method that we are applying: if both curvature and torsion are strong, then the interaction is also strong, and the WKB approximation fails.

\begin{figure}[H]
\begin{subfigure}{.5\linewidth}
\centering
\includegraphics[width=0.9\linewidth]{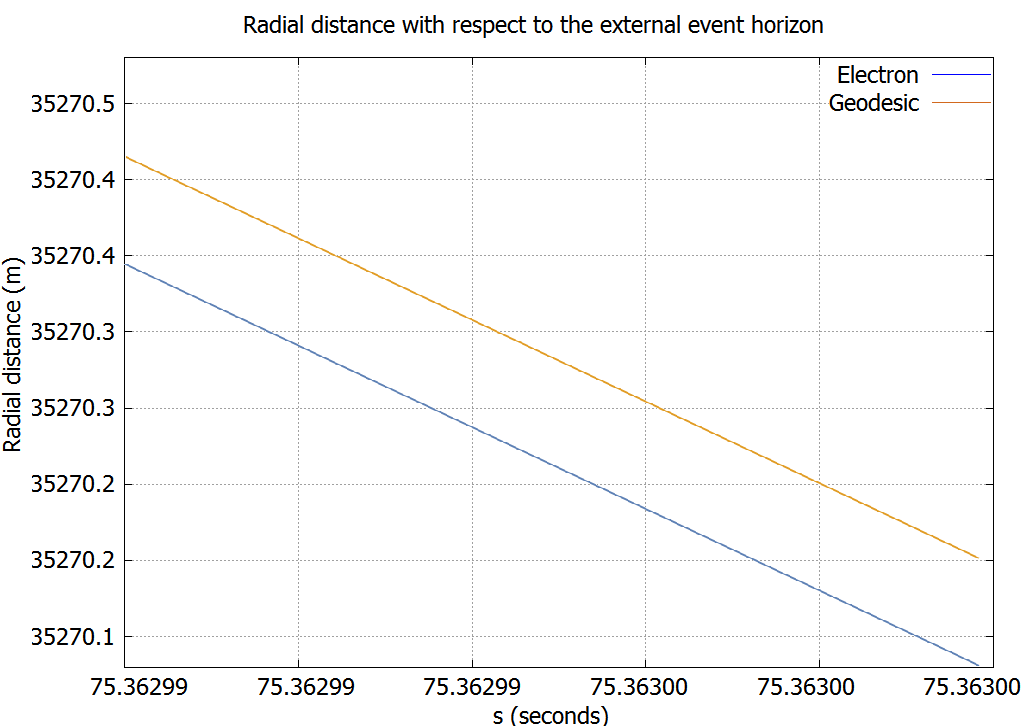}
\caption{}
\end{subfigure}%
\begin{subfigure}{.5\linewidth}
\centering
\includegraphics[width=0.9\linewidth]{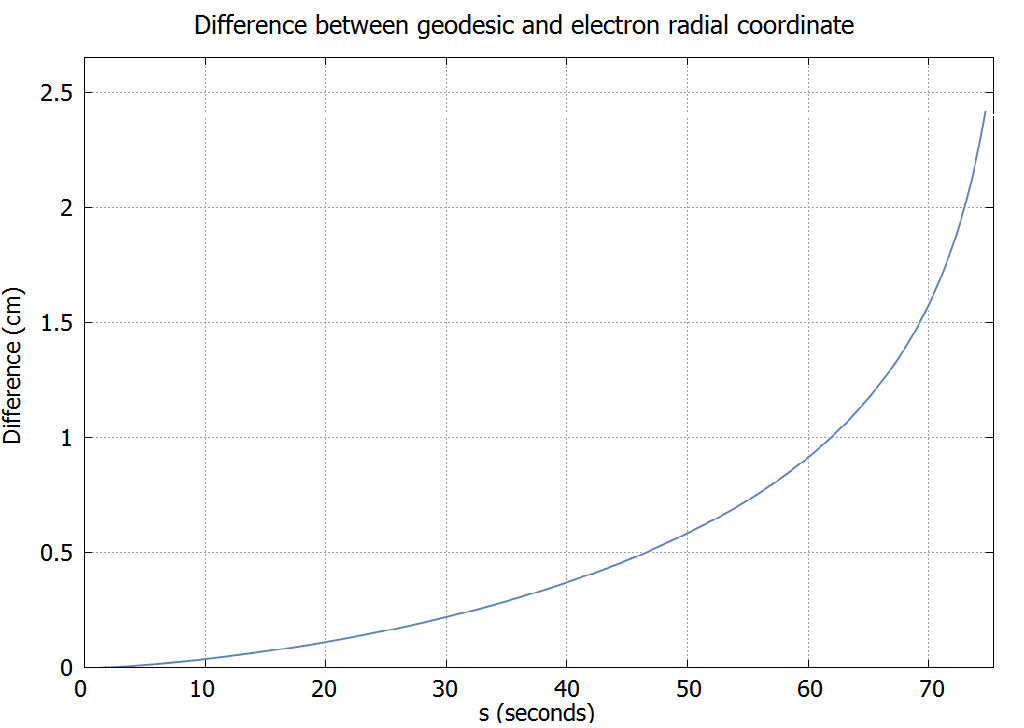}
\caption{}
\end{subfigure}
\caption{For this numerical computation, we have considered a BH with 24 solar masses and $\kappa=10$, with the electron located outside the external event horizon in the $\theta=\pi/2$ plane. We have assumed an electron with a radial velocity of 0.9 and a initial spin aligned in the $\varphi$ direction. (\textbf{a}) Trajectory at 35 km of the event horizon; (\textbf{b}) Relative position between the two particles.}
\label{fig:3}
\end{figure}

By taking into account these remarks, the main differences found for the present case study can be shown in Figure~\ref{fig:3}\footnote{Note that we have considered $d_{1}=1$, in order to simplify the reasoning.}. 
It is worthwhile to stress that any difference from the geodesic behavior in the radial coordinate would be an exclusive consequence of the torsion-spin coupling, with no presence of geometric terms provided by GR, by virtue of the dependence on $\kappa$ existing in the corresponding acceleration component. Indeed, it is possible to have situations under which the geodesic curves and the trajectories of spin 1/2 particles are distanced due to this effect, even by starting at the same point. Nevertheless, it is not strong enough to avoid their entrance to the BH region; hence, they present a singular behavior, as~expected. 

\section{Conclusions}

\label{conclusions}

In this manuscript, we have shown that the occurrence of singularities within strongly asymptotically-predictable spacetimes cannot be avoided by the introduction of the torsion tensor into their geometric structure. We have reviewed the singularity theorem that predicts the incompleteness of the trajectories of those spinning particles coupled to torsion and presented an explicit solution of a BH region in accordance with this theorem. Indeed, this configuration allows us to compute the trajectory of a Dirac fermion to conclude its singular behavior.

The physical effects derived by this post-Riemannian solution depend on the value of both the spin charge associated with the source of torsion and the coupling constant that determines the fundamental
strength of the interaction. By virtue of the purely quantum nature of the intrinsic angular momentum of matter, significant effects are expected only around extreme gravitational systems, such as neutron stars or specific BHs characterized by intense magnetic fields and sufficiently-oriented elementary spins.

In this sense, the design of experiments for the measurement of such implications related to the possible existence of a spacetime torsion is especially desirable and deserves further investigation in future works.

\vspace{6pt}

\authorcontributions{The three authors have contributed to the entire work. }

\funding{This work was partly supported by the projects FIS2014-52837-P (Spanish MINECO), FIS2016-78859-P (AEI
/FEDER, UE), Consolider-Ingenio MULTIDARK CSD2009-00064. FJMT acknowledges financial support from the National Research Foundation Grants 99077 2016--2018 (Ref. No. CSUR150628121624), 110966 (Ref. No. BS170509230233), the NRF IPRR
 (Ref. No. IFR170131220846), the Erasmus+ KA107 Alliance4Universities programme
and from the Van Swinderen Institute at the University of Groningen. }

\conflictsofinterest{The authors declare no conflict of interest.} 

\reftitle{References}

\end{document}